\documentclass[draft]{agujournal2019}
\usepackage{url} 
\usepackage{lineno}
\usepackage[inline]{trackchanges} 
\usepackage{soul}
\usepackage{color}
\draftfalse

\journalname{Geophysical Research Letters}

\begin{document}

%
%


\title{Resurgence of Lunar Volcanism: Role of Localized Radioactive Enrichment in a Numerical Model of Magmatism and Mantle Convection}
%
%




\authors{Ken'yo Uh\affil{1}, Masanori Kameyama\affil{2,3}, Gaku Nishiyama\affil{4,5,6}, Takehiro Miyagoshi\affil{3}, and Masaki Ogawa\affil{1}}

 \affiliation{1}{Department of Earth Sciences and Astronomy, The University of Tokyo, Komaba, Meguro, Japan}
 \affiliation{2}{Geodynamics Research Center, Ehime University, Matsuyama, Japan}
  \affiliation{3}{Japan Agency for Marine-Earth Science and Technology, Yokohama, Japan}
 \affiliation{4}{Institute of Planetary Research, German Aerospace Center, Berlin, Germany}
\affiliation{5}{Department of Earth and Planetary Science, The University of Tokyo, Hongo, Bunkyo, Japan}
 \affiliation{6}{RISE project, National Astronomical Observatory of Japan, Osawa, Mitaka, Japan}





\correspondingauthor{Ken'yo Uh}{u-kenyo0822@g.ecc.u-tokyo.ac.jp}



\begin{keypoints}
\item To understand the localized long-lasting volcanism of the Moon, we developed a numerical model of magmatism in the convecting mantle
\item The calculated volcanic activity has two peaks caused by ascent of partially molten plumes from the deep mantle to the uppermost level
\item Localized radioactive enrichment in the uppermost mantle plays an important role in the long-lasting volcanism with a couple of peaks
\end{keypoints}

%
%

%
%


\begin{abstract}
We develop a 2-D numerical model of magmatism and mantle convection to understand the volcanism on the Procellarum KREEP terrane (PKT) of the Moon, which continued for billions of years with two peaks of activities at 3.5-4 Gyr ago and around 2 Gyr ago. In our model, the effects of the PKT on lunar evolution are considered by initially imposing a region of localized radioactive enrichment. The calculated volcanism has two peaks induced by different mechanisms. The first peak occurs at 3.5-4 Gyr ago when magma generated in the deep mantle by internal heating ascends to the surface as partially molten plumes. The basaltic blocks in the uppermost mantle formed by this magmatism, then, sink to the deep mantle, triggering further plumes that cause the resurgence of volcanism at $\sim$2 Gyr ago. Our model shows that localized radioactive enrichment is important for the long-lasting volcanism with a couple of peaks.
\end{abstract}


\section*{Plain Language Summary}
Geological observations of the Moon have revealed that volcanism continued for billions of years in the Procellarum KREEP Terrane (PKT), a region enriched in radioactive heat-producing elements (HPEs); this long-lasting localized volcanism had peaks at 3-4 Gyr and around 2 Gyr ago. However, it remains unclear how volcanism has been maintained for so long. Here, we calculated a 2-D model of magmatism and mantle convection, considering a localized radioactive enrichment (EA) beneath the crust to model the PKT. Our simulations show that the first and second peaks of localized volcanism were induced by the ascent of partially molten plumes, but driven by different mechanisms. The first peak of the volcanism is accounted for by ascending partially molten plumes that are generated by strong internal heating at the base of the mantle. The second peak of the volcanism is driven by the descent of basaltic blocks formed by the earlier magmatism; the compositionally dense basaltic blocks founder into the deep mantle, triggering further partially molten plumes beneath the EA from 1.2 to 2.8 Gyr. Our model shows that localized radioactive enrichment in the uppermost mantle plays a critical role in the long-lasting volcanism of the Moon.

\section{Introduction}

Understanding the volcanism of the Moon has been a long-standing issue in studies of lunar mantle evolution \cite<e.g.,>{s&c1976,kirk&s,arai,Head2023rev}. Geological observations have revealed that volcanism has been particularly active in the Procellarum KREEP Terrane (PKT) located on the nearside, which is enriched in radioactive heat-producing elements (HPEs) \cite<e.g.,>{Jolliff,Lawrence,PKT,kamata2013}. Volcanism has occurred in other regions but has been less active \cite<e.g.,>{Shearer}. In the PKT, volcanic activity continued for several billion of years with a peak at 3.5-4 Gyr ago \cite<e.g.,>{hiesinger2000,hiesinger2003,hurwitz2013,whitten&head,nagaoka,Che2021,su2022}. 
This period of active volcanism coincides with the time of the radial expansion of the Moon: the Moon expanded globally by 0.5–5 km until around 3.8 Gyr ago \cite<e.g.,>{hana2013,hana2014,liang&Andrews,sawada} and then contracted globally over time \cite{Yue,Frueh}, by around 1 km or less in the past 1 Gyr \cite<e.g.,>{watters,watters2015,Klimczak,clark}.
The volcanic activity in the PKT, however, did not decline monotonically after the peak; it revived at around 2 Gyr ago \cite<e.g.,>{morota2011,cho2012,kato2017,giguere2022}. To understand the long-lasting history of lunar volcanism with two peaks in its activity, which mostly occurs in the PKT, we used a 2-D polar rectangular model of magmatism in the convecting lunar mantle \cite{u2023}.

Various numerical models have been developed to account for the long-lasting volcanism of the Moon \cite<e.g.,>{Solomon&head,WH2003,Breuer}. Some numerical models, in which the surface is covered by the HPE-enriched crust or a regolith layer as a blanket layer, suggest that magma persists in the uppermost mantle for 1-2 Gyr \cite<e.g.,>{k&s,Spohn,zie}. This partially molten region, however, extends globally and is unlikely to have caused the localized volcanism at the PKT. Some earlier researchers suggest that the localized volcanism is caused by one or some hot plumes growing from a layer of ilmenite-bearing cumulates (IBC) enriched in HPEs that developed on the core-mantle boundary (CMB) in the lunar early history \cite<e.g.,>{stegman,deVries,zhang2013a,zhang2017,zhang2022,zhang2023W} owing to the magma ocean and mantle overturn \cite<e.g.,>{hess&P,R&K}.
Although the plume enriched in the IBC component ascends by thermal buoyancy in these models, it is unclear if the large excess density of the IBC component, as expected from earlier models of mantle overturn, allows the ascent of such a compositionally dense plume in the Moon \cite<e.g.,>{Bars,li2019,yu,zhao2019}.
On the other hand, \citeA{PKT} and \citeA{Laneuville2013,Laneuville2018} assumed a locally HPE-enriched area at the top of the mantle in the initial condition and found that the area remained partially molten for more than 3 Gyr. 
In these models, however, the extraction of HPEs from the partially molten area by segregating magma \cite{cassen1973,cassen1979} is neglected, and HPEs remain in the uppermost mantle throughout the calculated 3 Gyr. \citeA{ogawa2018B} found that partially molten regions solidify within 2 Gyr when HPE-extraction by magmatism is considered.

To understand the mantle evolution of the Moon, constrained by its volcanic history as well as radial expansion/contraction history, we extend the 2-D model of magmatism and mantle convection that we have developed \cite{u2023}. In this model, magma is generated by decompression melting and internal heating, then migrates upward in partially molten regions as a permeable flow through the coexisting matrix; thermal, compositional, and melt buoyancy drive mantle convection in a Newtonian fluid with temperature-dependent viscosity; heat, mass, and incompatible HPEs are transported by magma. 
In our earlier study, we initiated the calculation with a compositionally stratified mantle, enriched with the IBC component at the base of the mantle \cite{u2023} following earlier studies of the magma ocean: during the last stage of crystallization of the lunar magma ocean, a dense IBC layer enriched in KREEP (K, rare earth elements, and P-rich material) develops at the top of the mantle and subsequently sinks to the CMB, a process called mantle overturn \cite<e.g.,>{alley,hess&P,bouk}.
In addition to this initial stratification, we also consider a localized enrichment of HPEs beneath the crust to model the PKT and investigate whether this enriched area (EA) induces a localized volcanism similar to that observed at the PKT (Figure \ref{initial}a).

\section{Model descriptions}
\label{Model description}

\begin{figure}
\noindent\includegraphics[width=\textwidth]{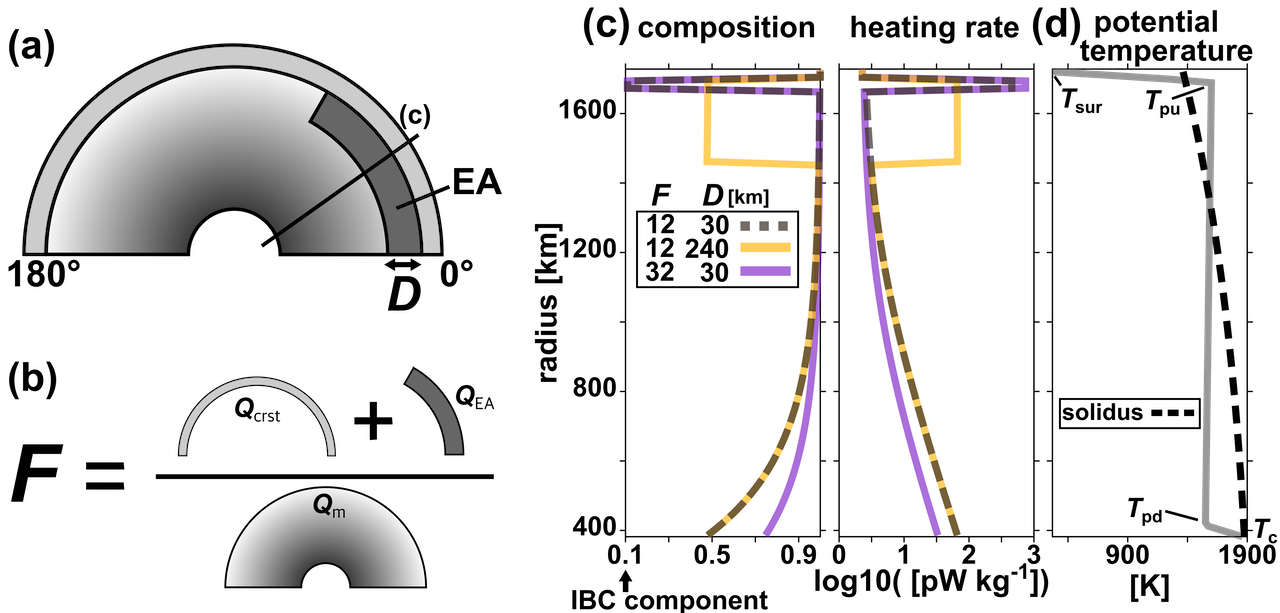}
\caption{ (a, b) An illustration of the initial distribution of heat-producing elements (HPEs). The distribution is laterally uniform except for the enriched area `EA’. In (b) $F$ is the ratio of the total amount of HPEs in the crust $Q_\mathrm{crst}$ and the EA $Q_\mathrm{EA}$ to that in the mantle $Q_\mathrm{m}$ (see Section S1-2 in Supporting Information). Also shown are (c) the initial distribution of the composition and heating rate along the line in (a); (d) the initial distribution of the potential temperature ($T_\mathrm{sur}$ = 270 K, $T_\mathrm{pu}$ = 1600 K, $T_\mathrm{pd}$ = 1550 K, and $T_\mathrm{c}$ = 1875 K are assumed in this figure).}
\label{initial}
\end{figure}

We used the 2-D annular model of mantle convection and magmatism developed in \citeA{u2023}. The finite difference numerical code calculates the energy, mass, and momentum equations of mantle convection in $R=\left[ \left( r, \theta \right) | \; 385 \; \mathrm{km} \leq r \leq 1735 \; \mathrm{km}, \; 0 \leq \theta \leq \pi \right]$ on a mesh with 128 (radial direction) times 256 (lateral direction) mesh points, under the Boussinesq approximation. (See Section S1 of Supporting Information for the detail of the basic equations.) To model the blanket effect of the crust \cite{zie}, we reduced the thermal diffusivity by a factor of two in a blanket layer 35 km thick at the top of the mantle.
The core is modeled as a heat bath with a uniform temperature, and its heat capacity is 0.01 times that of the mantle \cite{Kameyama}. The vertical sidewalls are insulated, while the surface boundary is fixed at 270 K; all boundaries are shear stress-free and impermeable to both magma and matrix.

The convecting material is a binary eutectic system composed of an olivine-rich material with a density of 3300 kg $\mathrm{m}^{-3}$ and an IBC material with a density of 3745 kg $\mathrm{m}^{-3}$ \cite<e.g.,>{snyder,Elkins-Tanton,liang2024}. The content of the olivine-rich component is denoted by $\xi$; the eutectic composition is at $\xi_\mathrm{e}$ = 0.1, corresponding to the basaltic composition. The bulk composition $\xi$ is calculated from the composition of the solid phase $\xi_{\mathrm{s}}$ and that of the liquid phase $\xi_{\mathrm{l}}$ written as
\begin{equation}
    \label{rho_s}
     \xi =\left( 1-\phi\right) \xi_{\mathrm{s}}  + \phi\xi_{\mathrm{l}} \,,
   \end{equation}
where $\phi$ is the melt-content. 
The solidus and liquidus temperatures are calculated from the phase diagram assumed for the binary eutectic system (see Appendix A in \citeA{u2023}). The convecting material contains HPEs that decay with time. 
Magmatism is modeled by the generation and upward migration of magma as a permeable flow \cite<e.g.,>{Mc1984,katz2008,miller}.
The flow is driven by the buoyancy of magma, and the relative velocity between the velocity of melt $\mathbf{u}$ and that of matrix $\mathbf{U}$ is calculated from
    \begin{equation}
    \label{M_num}
    \mathbf{u^*}-\mathbf{U^*} =
       -M g^*\frac{\phi^2}{\phi_0^3} \left( \rho_{\mathrm{s}}^* - \rho_{\mathrm{l}}^* \right) \, \mathbf{e_r}
       \,,
   \end{equation}
   where the asterisks stand for normalized quantities. $\rho_{\mathrm{s}} - \rho_{\mathrm{l}}$ is the density difference between solid and melt phases calculated as
   \begin{equation}
        \label{rho_l}
      \rho_{\mathrm{s}}^* - \rho_{\mathrm{l}}^* =  \beta \left(\xi_{\mathrm{l}}-\xi_{\mathrm{s}} \right) + \frac{\Delta v_{\mathrm{l}}}{v_{\mathrm{0}}}\left[ 1+
      \beta\left(1-\xi_{\mathrm{l}}\right)\right]\,
   \end{equation}
   (See Eqs. S6, S7 in Supporting information).
   Here, $g$ is the gravitational acceleration; $\phi_0$ = 0.05 is the reference melt-content; $\Delta v_{\mathrm{l}}/{v_{\mathrm{0}}}$ is the amount of density reduction by melting described in Eq. 5 of \citeA{u2023}.
      We assumed the reference permeability $M^*$ as
      \begin{equation}
        \label{Mpermf}
      M^* \equiv \frac{k_{\mathrm{\phi_{\mathrm{0}}}} \rho_0 {g_0} L}{\kappa \eta_{\mathrm{melt}}} \simeq 100 \,,
   \end{equation}
   where the parameter values are listed in Table S1.
   $\beta$ = 0.135 is a constant that expresses the sensitivity of density to the bulk composition. 
   In most cases, we assumed that the crustal density is higher than the density of magma for simplicity. However, the density of the lunar crust is lower than that of basaltic magma \cite<e.g.,>{morota2009,Taguchi,WH,HW,H&W2020}, and \citeA{Lourenco} suggest that the thermal history of the Moon can substantially depend on the crustal density inversion. 
   To clarify the effect of the density inversion, we increased the value of $\beta$ to 0.253 in the blanket layer in a case (Case crst-F12-D30): the crustal density is 2550 kg $\mathrm{m^{-3}}$ \cite{wieczorek2013} which is lower than the density of the basaltic magma at this value of $\beta$.

To simulate the localized enrichment of HPEs in the mantle beneath the PKT, we assumed an enriched area (EA) in 
$ \left[ \left( r, \theta \right) | \; 1700-D \; \mathrm{km} \leq r \leq 1700 \; \mathrm{km}, \; 0  \leq  \theta \leq 1/3 \,\pi \right]$ that is enriched in HPEs and the IBC component, as illustrated in Figure \ref{initial}a.
We examined the dependence of calculated volcanism on the thickness of the EA ‘$D$’ and on the ratio of the total amount of HPEs in the crust and the EA to the mantle ‘$F$’ (Figure \ref{initial}a, b); note that HPE-content in the crust is constant for simplicity (see Section S1-2 in Supporting Information).
The searched range of ‘$D$’ (10-240 km) is estimated from the thickness of the KREEP layer that remains after the mantle overturn \cite{Laneuville2013,Laneuville2018,Charlier,Schwinger,zhang2022nature,jones}.
We also take the searched range of ‘$F$’ (4-32) from the estimates in \citeA{Spohn}.


In the initial condition, we assumed that the mantle is compositionally stratified to simulate the layering of the lunar mantle inferred from previous studies of the magma ocean and mantle overturn \cite<e.g.,>{Parmentier2002,hess&P}. The contents of HPEs and the IBC component increase exponentially with depth as illustrated in Figure \ref{initial}c. The average of the internal heating rate over the entire mantle is $q_0 = 14.7 $ $\mathrm{pW \, kg^{-1}}$ at 4.4 Gyr estimated from Table S2 in Section S1-1.
The initial temperature distribution (Figure \ref{initial}d) is also taken from earlier models of the lunar mantle overturn \cite<e.g.,>{bouk,yu,zhao2019} and the evolution model \cite{u2022}. The details of the initial condition are described in Section S1-2. 

  \section{Results} 
  
  \label{results}

\begin{figure}
  \noindent\includegraphics[width=\textwidth]{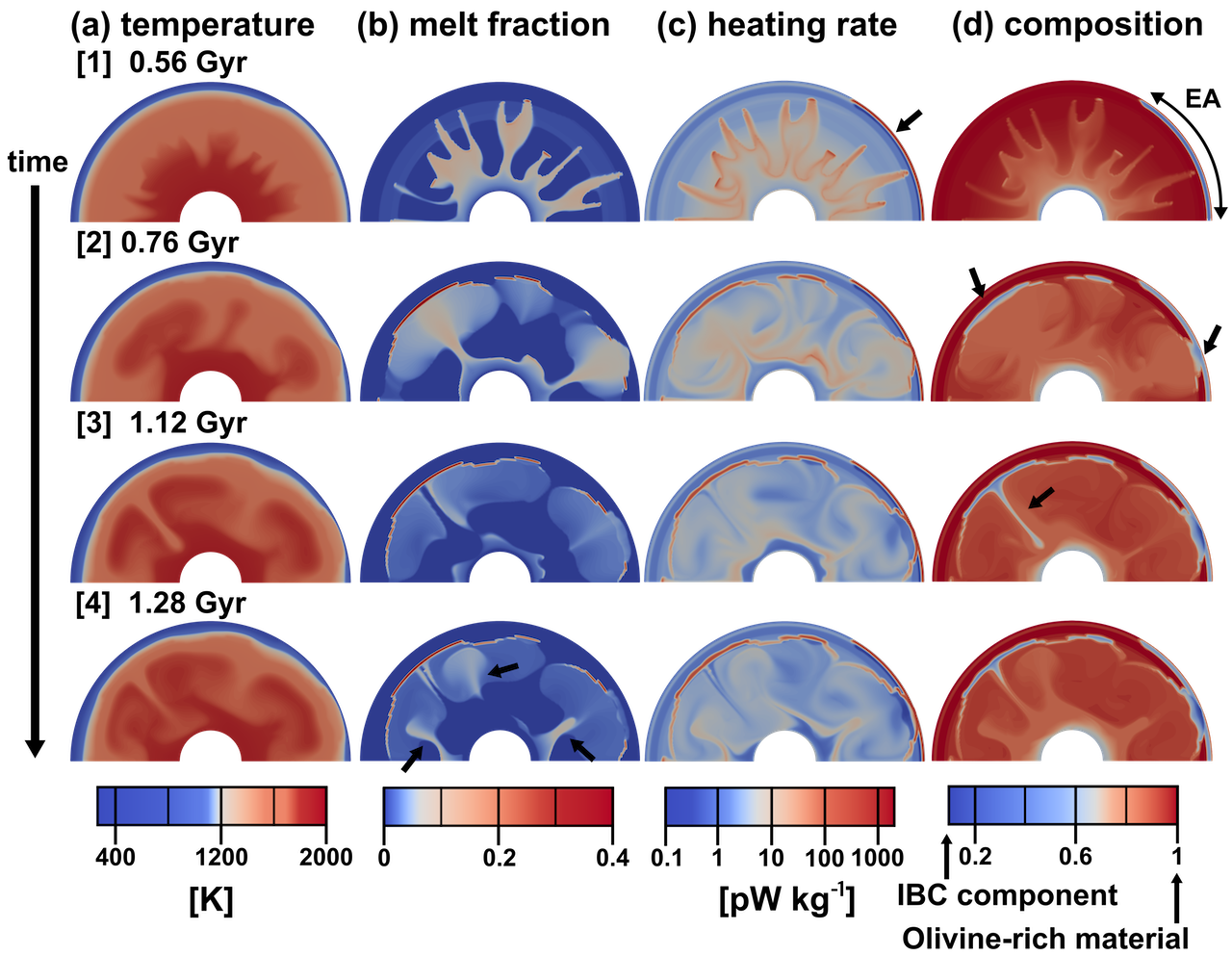}
 \caption{Snapshots of the distributions of (a) temperature, (b) melt-content, (c) internal heating rate, and (d) bulk composition calculated for the reference case ($F = 12$ and $D = 30$ km; Case F12-D30). The elapsed times are indicated in the figure. In (d), the blue color stands for the basaltic composition enriched in the IBC component, while the red color stands for the olivine-rich end-member. The numbers [1] to [4] correspond to those of Figure \ref{distribution}b.}
  \label{for_journal}
 \end{figure}

\begin{figure}
\noindent\includegraphics[width=\textwidth]{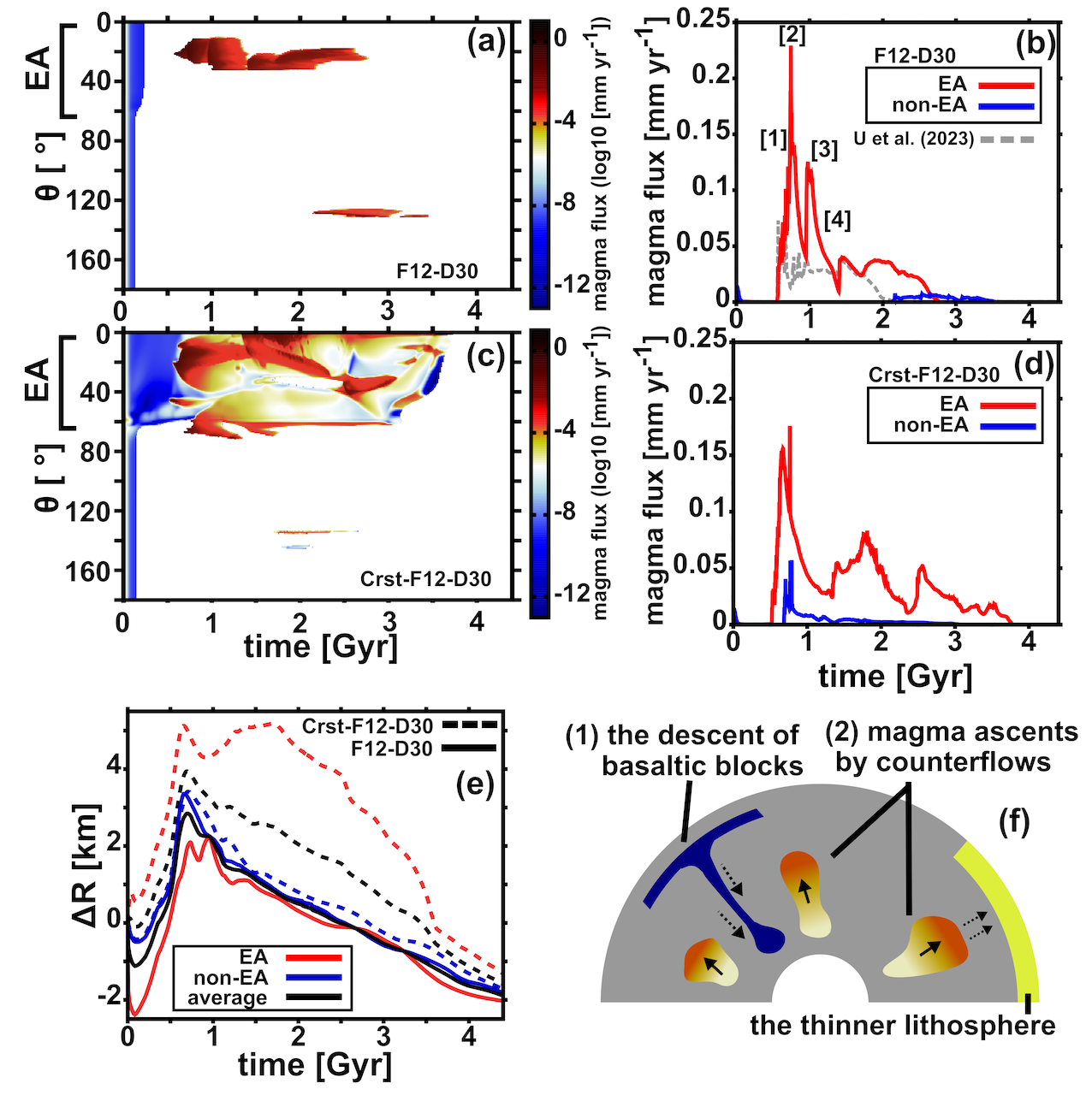}
\caption{(a) The distribution of the upward flux of magma that passes upward through the depth level of 100 km and (b) the average flux of magma around the EA (the red line) and outside the EA (the blue line) at the depth level of 100 km in Case F12-D30 (the reference case). Also shown are (c \& d) the magma flux calculated in Case crst-F12-D30 where the crust is less dense than basaltic magma; (e) the history of radius change $\Delta R$ calculated by Eq. S17 in Section S1-3; (f) a schematic illustration of long-lasting volcanism with a couple of peaks in the PKT calculated in the model.
In (b, d, and e), the blue lines and red lines stand for the average values in $ \left[0  \leq  \theta \leq 1/3 \,\pi\right]$ (beneath the EA) and in $ \left[1/3 \,\pi < \theta \leq \pi \right]$ (outside the EA), respectively, while the black lines stand for the total average values in $ \left[0  \leq \theta \leq \pi  \right]$.
In (b), the gray line shows the magma flux in our earlier model \cite{u2023} where the EA is not considered (other parameters are the same as those assumed in Case F12-D30); the numbers [1]–[4] correspond to those of Figure \ref{for_journal}. }
\label{distribution}
\end{figure}

In Figures \ref{for_journal} and \ref{distribution} as well as the animation (Movie S1 in the supporting file), we present the evolution of the mantle calculated in Case F12-D30 where $F$ = 12 and $D$ = 30 km (see also figures in Supporting Information).

Magma is generated in the deep mantle owing to the strong internal heating assumed in the initial condition (Figure \ref{initial}c) and migrates upward as partially molten plumes to reach the uppermost mantle within the first 0.7 Gyr (Figures \ref{for_journal}b, c; see also Figure S1a); melt buoyancy is the main driving force of this flow. 
Magma rises to the depth level of 100 km beneath the EA in $ \left[0  \leq  \theta \leq 1/3 \,\pi\right]$, which is shallower than beneath the area outside the EA (Figures \ref{distribution}a, b). 
This is because the lithosphere is thinned beneath the EA owing to the strong heating by HPEs in the EA (Figure \ref{for_journal}a, c). 
The magma transports HPEs and the basaltic component to the uppermost mantle, forming basaltic blocks (see the arrows in Figure \ref{for_journal}d for 0.76 Gyr). A deeper part of the compositionally dense basaltic blocks eventually founders to the CMB as magma accumulates and the blocks grow (see the arrow in Figure \ref{for_journal}d for 1.12 Gyr).
We found that the core temperature drops by about 40 K in 100 Myr as the cold basaltic blocks descend to the CMB (see Figure S1b, c). 
The descending basaltic blocks trigger further ascent of partially molten plumes from the deep mantle driven by melt buoyancy, as depicted by the arrows in Figures \ref{for_journal}b. 
Accordingly, the upward magma flux beneath the EA becomes significant again from 1.2 to 2.8 Gyr
(Figure \ref{distribution}b). The resurgence of magma flux is not found in our earlier models that are devoid of the EA (see the gray dotted line in Figure \ref{distribution}b). In contrast, the magma flux is negligibly small throughout the calculated history outside the EA due to the thicker lithosphere (Figure \ref{distribution}a, b).
After 2.8 Gyr, the partially molten region in the mantle shrinks with time due to the decay of HPEs but persists until 4.4 Gyr (Movie S1 as well as Figure S1d). From further numerical experiments with various values of $D$ in the range of 10-240 km and $F$ in the range of 4-32, we found that the localized volcanism revives and continues for more than 1.5 Gyr in the cases with $D$ $<$ 120 km and 8 $<$ $F$ $<$ 24 (see Section S2-2 including Table S3 and Figure S2 in Supporting Information). 

The resurgence of magma flux beneath the EA observed in the reference case is more pronounced in Case crst-F12-D30 where the crustal density is lower than the density of the basaltic magma (see Figure \ref{distribution}c, d as well as Figure S3). The upward magma flux shown in Figure \ref{distribution}c is substantial beneath the entire EA until 3.7 Gyr, and magma enriched in HPEs remains along the crust-mantle boundary even at as late as 3.5 Gyr (Figure S3). The resurgence of magma flux is so significant in crst-F12-D30 because the lower density of the crust prevents magma from transporting HPEs to the surface; HPEs remain beneath the crust to keep the lithosphere beneath the EA thinner compared with that in the reference case.

We found that strong internal heating and compositionally high density in the EA, as well as initial mantle stratification, are important for the resurgence of magma flux beneath the EA (see Figure S4-S7).
The resurgence of magma flux is not observed in Case noHeat-F12-D30 where the enrichment of HPEs in the EA is not considered (Figures S4a and S5). The resurgence occurs only briefly in Case noBa-F12-D30 where the compositional high density in the EA is not considered (Figures S4b and S6).
This is due to the smaller density difference between the mantle and magma $\rho_{\mathrm{s}} - \rho_{\mathrm{l}}$ in the EA compared to the reference case, resulting in lower permeable flow there (Eqs. \ref{M_num} and \ref{rho_l}).
On the other hand, magma flux at the depth level of 100 km is not observed since 0.5 Gyr after the start of the calculation in Case noStra-F12-D30 where the HPEs and the composition $\xi$ are vertically uniform in the initial mantle (Figures S4c and S7). The deep mantle is not enriched in HPEs in this case, and sufficient magma generation does not occur to ascend to the uppermost mantle.

Figure \ref{distribution}e shows that the Moon expands globally by more than 3 km for the first 0.7 Gyr of the calculated history and then contracts at a rate of around -1 km $\mathrm{Gyr^{-1}}$ in the past 1 Gyr in the reference case. The decomposition of radius change $\Delta R$ into the component caused by melting/solidification and by thermal expansion/contraction shows that the expansion is mainly caused by melting, while the contraction is mainly caused by thermal contraction (Figure S8).
The overall feature of the radius change is not affected by the crustal density (see the black dotted line in Figure \ref{distribution}e). However, there is a drop in the radius beneath the EA $\Delta R_\mathrm{EA}$ at around 3.7 Gyr due to the solidification of the mantle in Case crst-F12-D30 (see Figure S9). This is because a partially molten region that persists until that time along the crust-mantle boundary beneath the EA solidifies (Figure \ref{distribution}c; see also Figure S3).


\section{Discussions} 
Figure \ref{distribution}f illustrates how an enriched area (EA) causes a long-lasting volcanism in our models.
 In the earlier stage of mantle evolution, magma is generated in the deep mantle by internal heating and ascends to the uppermost mantle as partially molten plumes. In the area beneath the EA, magma rises to a shallower depth than in the area outside the EA because the lithosphere there is thinner owing to the strong internal heating assumed in the initial condition. This magmatism results in the formation of basaltic blocks beneath the lithosphere. The basaltic blocks outside the EA eventually sink to the CMB, triggering further plume activity that causes to a second peak in volcanism.

\subsection{Comparison With Earlier Models of Lunar Volcanic History}
A comparison with earlier mantle convection models of lunar volcanic history, which mostly occurs in the PKT and its surrounding area, shows the crucial role of melt buoyancy in our models. Some models suggest that localized volcanism is caused by thermal plumes from an HPE-enriched IBC layer on the CMB \cite<e.g.,>{zhong,zhang2013b}. In these models, however, the compositional density contrast between the basal IBC layer and the overlying olivine-rich mantle is substantially less than 160 kg $\mathrm{m^{-3}}$ (see Figure 6 in \citeA{li2019}), the value suggested from some recent mantle overturn models \cite<e.g.,>{yu,zhao2019,maurice2024}. At this density contrast, the basal layer remains convectively stable and does not rise as a plume by thermal buoyancy alone \cite{Bars,deVries}.
In contrast, a large fraction of the IBC component in the deep mantle arises as partially molten plumes mostly driven by melt buoyancy in our models (Figure \ref{for_journal}) although the density contrast is more than 200 kg $\mathrm{m^{-3}}$ in our reference case (Figure \ref{initial}c).


Our models also suggest the reason why volcanism continues so long in the PKT. 
In some of earlier models, a locally HPE-enriched area at the top of the mantle assumed in the initial condition remains partially molten for more than 3 Gyr \cite{PKT,Laneuville2013,Laneuville2018}. 
HPE-extraction from the enriched area by magmatism is, however, neglected in these models as we discussed before.
In our models, magmatism continues long despite that HPE-transport by magma is taken into account, because the basaltic blocks formed by early magmatism descend to the CMB to trigger further ascent of partially molten plumes (Figures \ref{for_journal} and \ref{distribution}a, f).
Besides, the melting history of the uppermost mantle depends on the crustal density.
In Case crst-F12-D30 where the crustal density is lower than the density of the basaltic magma, HPEs are not extracted to the surface but remain at the base of the crust (see Figure S3). 
The HPEs at the crust-mantle boundary keep the lithosphere thin enough for revived plume activity to cause volcanism.
The resurgence of magma flux is so significant beneath the entire EA for more than 3.5 Gyr due to the plume activity (Figure \ref{distribution}c, d).
By comparing magma flux under varying initial conditions (Figures S4-S7), our simulations also show that both the initial mantle stratification and the EA beneath the crust play crucial roles in the resurgence of volcanism after 1 Gyr.
Note that a detailed discussion of volcanic history is possible in our models because we calculate the magma flux directly, rather than estimating it from the distribution of magma in the mantle, as is done in earlier models \cite<e.g.,>{Wood,solomon&T,k&s,u2022}.

\subsection{Comparison With the Observed Features of the Moon}
The magma flux beneath the EA shown in our reference case is consistent with the observed history of volcanism in the PKT. Magma generated at depth by strong internal heating rises (Figure \ref{for_journal} for 0.56 Gyr and 0.76 Gyr), causing the peak of volcanism at 3.5-4 Gyr ago \cite<e.g.,>{hiesinger2000,whitten&head}. The resurgence of volcanism triggered by the descending basaltic blocks is consistent with the observed volcanic history in the PKT with the second peak at around 2 Gyr ago \cite<e.g.,>{cho2012,tian2023}. 
The early expansion and later contraction of the planet in the reference case (Figure \ref{distribution}e) are also consistent with those suggested from gravity gradiometry data and tectonic features \cite<e.g.,>{hana2013,bogert,matsu}. 
The plumes triggered by descending basaltic blocks are depleted in HPEs compared to the plumes that ascend in the early history (see the frames of 0.76 Gyr and 1.28 Gyr in Figure \ref{for_journal}c, d; see also the animation). The plume activity in the later history can account for the HPE-depleted young basalts observed in the PKT \cite{Che2021,li2021,su2022}.
It is, however, difficult to account for the temporal trend of increasing Ti content in the HPE-depleted mare basalts in the PKT after around 2.3 Gyr ago \cite{kato2017,sato2017,zhang2022titanium}.
This trend may be due to the remelting of basaltic blocks in the uppermost mantle by the plumes that rise later (Figures \ref{for_journal}). 
To clearly understand this trend, it is necessary to carefully model magma evolution caused by fractional crystallization \cite<e.g.,>{luo2023}.

 
In all of our calculated models, the heat flux on the CMB $q_\mathrm{CMB}$ is too low to drive the long-lasting core dynamo in all of our calculated models. 
In the reference model, the heat flux is close to 0 mW $\, \mathrm{m^{-2}}$ since the first few hundred million years (see Figure S1c).
The same conclusion has been reached in many of previous mantle convection models \cite<e.g.,>{k&s,Laneuville2013,maurice2024}. A previous model proposes that $q_\mathrm{CMB}$ becomes temporally high enough to induce the core dynamo when the IBC-rich layer on the CMB is internally heated and rises as plumes by thermal buoyancy \cite{stegman}. Although hot plumes rise from the IBC-rich layer in our model too, a part of the layer remains on the CMB for more than 800 Myr to keep $q_\mathrm{CMB}$ around 0 mW $\, \mathrm{m^{-2}}$ (Figure \ref{for_journal}a, b for 0.76 Gyr; see also Figure S1b). $q_\mathrm{CMB}$ temporally increases only slightly at around 1.4 Gyr as cold basaltic blocks sink to the CMB (see Figure S1c). The flux is too small to drive the core dynamo \cite<e.g.,>{christensen2009,weiss&Tikoo}. 
It is likely that dynamo mechanisms other than secular cooling of the core are needed to sustain the long-lasting core dynamo throughout the history of the Moon, such as core crystallization \cite<e.g.,>{Laneuville2014,Evans2018}, a basal magma ocean \cite{Scheinberg2018,Hamid2023}, and precession of the rotation axis \cite<e.g.,>{dwyer2011,stys2020}.
It is also important to re-examine the history of magnetic fields on the Moon \cite<e.g.,>{tikoo2017,tarduno2021,jung2024}.

 Further calculations in a 3-D spherical mantle are important to more quantitatively predict the history of magma flux, radial expansion/contraction, and thermal history of the Moon and to compare the predictions with such observations as the temperature distribution in the Moon \cite<e.g.,>{khan2006,khan2014}. 
As already found by \citeA{Guerrero} for the Moon, 2-D polar rectangular models of mantle convection tend to predict higher average mantle temperatures compared to 3-D spherical models, especially when the core size is small.
In the reference case, the mid-mantle temperature at 4.4 Gyr is about 100 K higher than current lunar estimates (Figure S1d), which may result from the geometry of the convecting vessel.

\section{Conclusions}
 To understand the long-lasting localized volcanism in the PKT (the Procellarum KREEP terrane) with the first peak at 3-4 Gyr ago and the second one at about 2 Gyr ago, we numerically simulated a 2-D polar rectangular model of magmatism and mantle convection where a localized heat-producing elements (HPEs) enrichment beneath the crust, called enriched area (EA) is considered to model the PKT in the initial mantle stratification (Figure \ref{initial}). 

 Our simulations show that the first and second peaks of localized volcanism, occurring at about 1 Gyr intervals, were induced by ascent of partially molten plumes driven by different mechanisms.
The first peak of the volcanism is accounted for by magma ascent as partially molten plumes from the deep mantle induced by internal heating there (Figures \ref{for_journal} and \ref{distribution}b). 
Magma ascends to the uppermost mantle beneath the EA because the lithosphere there is thinned by the strong internal heating in the EA. 
The second peak of the volcanism, in contrast, is caused by counterflows driven by the descent of basaltic blocks formed by the earlier magmatism beneath the lithospheric lid; a deeper part of the compositionally dense basaltic blocks sinks into the deep mantle owing to their negative buoyancy, triggering further magma ascent beneath the EA from 1.2 to 2.8 Gyr (Figure \ref{distribution}b, f). 
We found that the mantle evolves as illustrated in Figure \ref{distribution}f when the thickness of the EA is $D$ $<$ 120 km and the ratio of the total amount of HPEs in the crust and the EA to the mantle is 8 $<$ $F$ $<$ 24 (see Table S3).
Our models suggest that material transport and melt buoyancy by magmatism play critical roles in the lunar mantle evolution. 
To reproduce the long-lasting volcanism with a couple of peaks in the PKT, localized radioactive enrichment in the uppermost mantle is important. 


\section*{Data Availability Statement}

The official version of the animations and Supporting Information as well as dataset will be made available if it has been accepted by GRL. However, the unofficial version of the animation and Figures S1-S9 can be accessed via the following links.

https://youtu.be/Mdhh9a7P-zY

https://youtu.be/m1dCnxtrIgE

\bibliography{agusample.bib}

\nocite{sakamaki}
\nocite{vankan}
\nocite{xu}
\nocite{mei}
\nocite{Scott}
\nocite{Hirth}
\nocite{dygert}
\nocite{karato2013}
\nocite{Siegler2022}
\nocite{Siegler2014}
\nocite{Geodynamics}
\nocite{garciaA}
\nocite{garciaB}
\nocite{McDonough&Sun}
\nocite{Lodders}
\nocite{Tayor&Wieczorek}
\nocite{ogawa2018}
\nocite{ogawa2020}

\acknowledgments
The authors would like to extend their sincere appreciation to M. Kayama at the University of Tokyo and T. Yanagisawa at JAMSTEC for their constructive comments. This work was supported by JST SPRING Grant Number JP-MJSP2108 and JSPS KAKENHI Grant Number JP202412823 of Japan. This work was also supported by: the Joint Usage/Research Center PRIUS at Ehime University, the Earth Simulator of Japan Agency for Marine-Earth Science and Technology (JAMSTEC), ``Exploratory Challenge on Post-K Computer’’ (Elucidation of the Birth of Exoplanets [Second Earth] and the Environmental Variations of Planets in the Solar System), and ``Program for Promoting Research on the Supercomputer Fugaku’’ (Toward a unified view of the universe: from large scale structures to planets). This work used computational resources of the supercomputer Fugaku provided by the RIKEN Center for Computational Science through the HPCI System Research Project (Project ID: hp230204, hp240219). Animations and some figures were drawn with the ParaView by Sandia National Laboratory, Kitware Inc., and Los Alamos National Laboratory.


%
%




%
%
%
%
%

\end{document}